\documentclass[journal,transmag,10pt]{IEEEtran}
\usepackage{amsmath}
\usepackage{amssymb}
\usepackage{graphicx}
\usepackage{float}
\usepackage{hyperref}
\usepackage[style=ieee, url=false, backend=biber,]{biblatex}
\usepackage{xcolor}
\usepackage{rotating}
\usepackage{pdflscape}
\usepackage{parskip}

\usepackage[symbol]{footmisc}
\graphicspath{ {./images/} }
\usepackage{subcaption}

\DeclareBibliographyDriver{article}{%
  \printnames{author} % Print the authors' names
  \newunit\newblock
  \printfield{title} % Print the title of the article
  \newunit\newblock
  \printfield{shortjournal}
  \setunit{\addspace} % Add space between journal title and volume
  \printfield{volume} % Print the volume number
  \setunit*{\addcomma\space} % Add comma and space before pages
  \printfield{pages} % Print page numbers
  \setunit*{\addcomma\space} % Add comma and space before year
  \usebibmacro{date} % Use macro to print date/year 
  \finentry
}

\AtEveryBibitem{%
  \clearname{translator}
  \clearname{number}
  \clearlist{number}
  \clearfield{file}
  \clearfield {pubstate}
  \clearfield {url}
  \clearfield {urldate}
  \clearfield{urlyear}
  \clearfield{urlmonth}
  \clearfield{doi}
  \clearfield{archiveprefix}
  \clearfield{keywords}
  \clearfield{eprint}
}

\begin{document}
\title{LoC-LIC: Low Complexity Learned Image Coding Using Hierarchical Feature Transforms}

\author{\IEEEauthorblockN{Ayman A. Ameen\IEEEauthorrefmark{1} \IEEEauthorrefmark{2},
Thomas Richter\IEEEauthorrefmark{1}, and
André Kaup\IEEEauthorrefmark{3},}

% Ayman Ameen  https://orcid.org/0009-0008-0752-2406

\IEEEauthorblockA{\IEEEauthorrefmark{1}Fraunhofer Institute for Integrated Circuits IIS, Erlangen, Germany}

\IEEEauthorblockA{\IEEEauthorrefmark{2} Department of Physics, Faculty of Science, Sohag University, Egypt}

\IEEEauthorblockA{\IEEEauthorrefmark{3} Friedrich-Alexander University at Erlangen-Nürnberg, Erlangen, Germany}}
\maketitle
\begin{abstract}
Current learned image compression models typically exhibit high complexity, which demands significant computational resources. To overcome these challenges, we propose an innovative approach that employs hierarchical feature extraction transforms to significantly reduce complexity while preserving bit rate reduction efficiency. Our novel architecture achieves this by using fewer channels for high spatial resolution inputs/feature maps. On the other hand, feature maps with a large number of channels have reduced spatial dimensions, thereby cutting down on computational load without sacrificing performance. This strategy effectively reduces the forward pass complexity from \(1256 \, \text{kMAC/Pixel}\) to just \(270 \, \text{kMAC/Pixel}\). As a result, the reduced complexity model can open the way for learned image compression models to operate efficiently across various devices and pave the way for the development of new architectures in image compression technology.

\end{abstract}

\section{Introduction}

Recently, Learned image compression models have achieved significant gains in bit rate reduction; however, traditional image and video compression methods, such as JPEG 2000, JPEG XL, HEVC \cite{sullivanOverviewHighEfficiency2012} and VVC \cite{brossDevelopmentsInternationalVideo2021}, still largely dominate the field. Despite the potential benefits of the learned compression methods, the adaptation of these methods has been slow. One main reason is their high complexity and considerable resource requirements. 

Most learned image compression architectures use transform analysis and synthesis with convolutional layers, maintaining a fixed number of channels and resizing the input for each (residual) layer. However, the total multiply-accumulate operations (MACs) per pixel increase with larger input sizes. Using a fixed number of channels results in higher MACs in initial layers without providing substantial benefits. To address this, we propose a hierarchical feature extraction approach with lower complexity by employing fewer channels for larger feature maps and more channels for smaller feature maps.

Our novel approach utilizes hierarchical feature extraction transforms to map images from the pixel domain to the latent domain and vice versa, reducing both memory and computational complexity. The key features of our approach include:

\begin{itemize}
    \item Low complexity autoencoder through our novel hierarchical feature extraction, which has progressively deeper feature representations with a lower number of feature maps for larger sizes and higher features for smaller sizes, allowing reduction forward pass complexity from 1256 kMAC/Pixel to only 270 kMAC/Pixel.
    \item Hyper-autoencoder with multi-reference entropy model maintaining competitive performance to the state-of-the-art models. 
    \item A large dataset that spans the large part of the image space manifold.
\end{itemize}

\begin{figure}[h]
    \centering
    \includegraphics[width=0.5\textwidth]{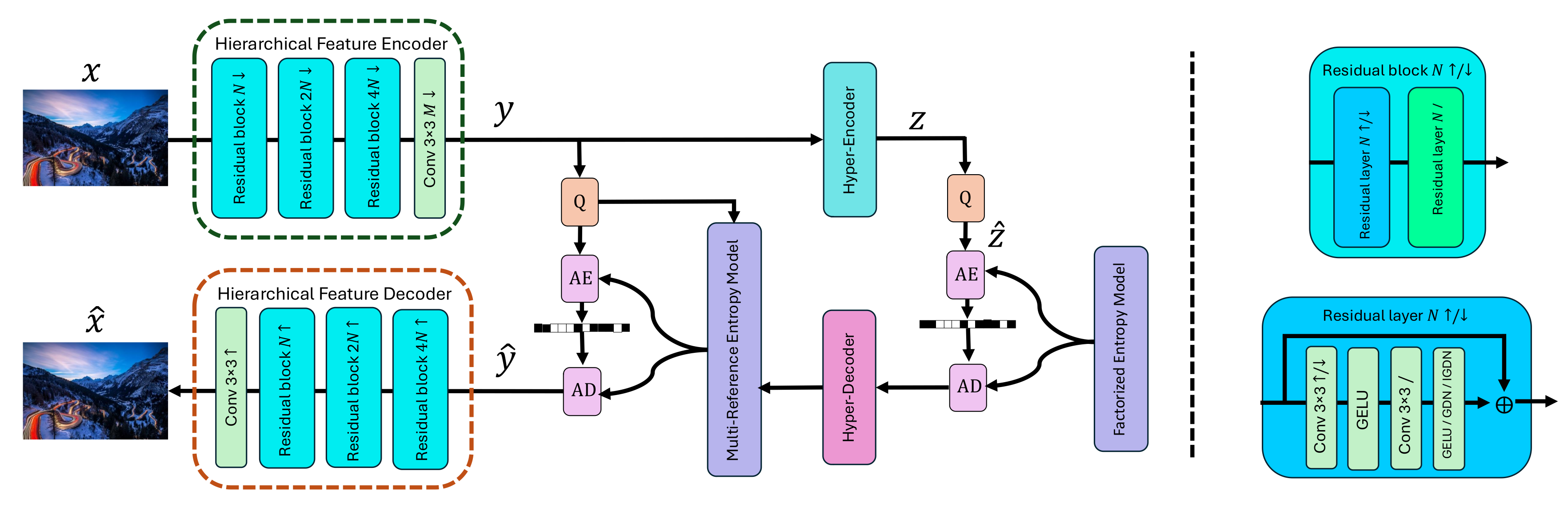}
    \caption{Overall view of the proposed architecture with hierarchical feature encoder and decoder.}
    \label{fig:model}
\end{figure}
\section{Related Works}

Traditional image compression algorithms, such as JPEG 2000, often lack the flexibility of non-linear mapping to input data through a learning process. To address these limitations, a novel learned image compression method using autoencoders has been developed. This approach involves training an autoencoder on large datasets of images or videos \cite{jiangMLICLinearComplexity2024, jiangMLICMultiReferenceEntropy2023, heELICEfficientLearned2022, theisLossyImageCompression2017}. 

The key advantage of the learned image codec lies in the autoencoder's capacity to map images into a low-entropy, high-dimensional latent space, and subsequently reconstruct them back into the image space in a lossy manner \cite{balleEndtoendOptimizedImage2017, balleVariationalImageCompression2018}. The learned image codec network architecture is typically composed of analysis and synthesis transform functions, which can be implemented using pure convolutional layers with a fixed number of channels $N$, \cite{liuComprehensiveBenchmarkSingle2020}. Some architectures incorporate residual connections with convolutional layers as the base model \cite{liuLearnedImageCompression2023}, while others utilize transformer-based layers \cite{luTransformerbasedImageCompression2022} or combine attention mechanisms with convolutional layers for enhanced performance \cite{chengLearnedImageCompression2020}. The goal is for the reconstructed image $\hat{x}$ to closely resemble the original image $x$ while minimizing the bit-rate $R$ used for the latent representation. Generally, higher entropy results in lower distortion and vice versa. Therefore, optimization aims to balance distortion $D(x,\hat{x})$ with entropy measured in bitrate $R$ (bits per pixel).  A Lagrangian multiplier is employed to manage this trade-off between distortion and target bit-rate \cite{fuLearnedImageCompression2023, chengLearnedImageCompression2020}. Quantization plays a critical role in bitrate reduction but introduces non-differentiability, hindering its direct integration with gradient-based optimization. To address this, continuous relaxations of the quantization operator, such as the straight-through estimator (STE), are widely adopted to enable differentiable approximations during training. An alternative strategy involves replacing deterministic quantization with stochastic rounding, which introduces non-biased gradients and has proven effective in recent compression frameworks \cite{minnenJointAutoregressiveHierarchical2018,balleVariationalImageCompression2018}. Both approaches circumvent the discontinuity in backpropagation while preserving quantization benefits. Stochastic rounding can be easily implemented by adding uniform noise to the unquantized values. A hybrid approach that combines both straight-through estimation and stochastic rounding also exists \cite{jiangMLICLinearComplexity2024}.  Utilizing context information allows for more efficient data compression by reducing the bit-rate necessary for encoding. Context-based entropy models exploit surrounding or neighboring information to better predict and compress the current data. This strategy is particularly important in neural image compression, as it enables accurate bit-rate estimation while minimizing redundancy.

To enhance compression efficiency, various context-based entropy models have been proposed. An autoregressive model was introduced to condition each pixel on previously decoded pixels for more effective context modeling \cite{minnenJointAutoregressiveHierarchical2018}. Another approach is the checkerboard convolution, which divides the latent representation into anchor and non-anchor parts, using the anchor part to extract context for the non-anchor part \cite{chengLearnedImageCompression2020}. Furthermore, channel-wise context models \cite{liuUnifiedEndEndFramework2020}, and channel-wise models with unevenly grouped contexts \cite{heELICEfficientLearned2022}, have been developed to exploit redundancy between channels. Recently, an attention-based architecture has been proposed to capture a diverse range of correlations within the latent representation \cite{jiangMLICMultiReferenceEntropy2023,jiangMLICLinearComplexity2024}.

Another promising approach for learned image compression involves using an overfitted neural network to represent image data as a continuous neural function instead of discrete pixel values. This neural function can be evaluated to reconstruct the RGB values of image pixels. Various efforts have been made to represent entire datasets, such as MNIST, using neural functions for resolution-agnostic representations \cite{garneloConditionalNeuralProcesses2018, kimAttentiveNeuralProcesses2019, gordonConvolutionalConditionalNeural2020}. A significant advantage of modeling images as neural functions is their resolution agnosticism: images are represented continuously and can be evaluated at any desired resolution. This approach assumes that image signals are inherently continuous.

The COIN framework \cite{dupontCOINCOmpressionImplicit2021} introduced the concept of using overfitted learnable functions for image compression. It utilizes a straightforward multilayer perceptron (MLP) to map pixel coordinates to their respective $RGB$ values by effectively using periodic activation functions \cite{sitzmannImplicitNeuralRepresentations2020}. While COIN's performance was on par with JPEG compression, it was constrained by its inability to take advantage of pixel locality due to the inherently non-local characteristics of MLPs. This issue was addressed by employing a multi-resolution latent representation followed by a non-linear MLP \cite{mullerInstantNeuralGraphics2022}. COOL-CHIC \cite{laduneCOOLCHICCoordinatebasedLow2023, leeEntropyConstrainedImplicitNeural2023, leguayLowComplexityOverfittedNeural2023, blardOverfittedImageCoding2024} introduced an advanced overfitted learned image codec with reduced decoding complexity, which significantly improved compression efficiency compared to COIN.
\begin{figure*}[h]
    \centering
    \begin{subfigure}{0.45\textwidth}
        \centering
        \includegraphics[width=\textwidth]{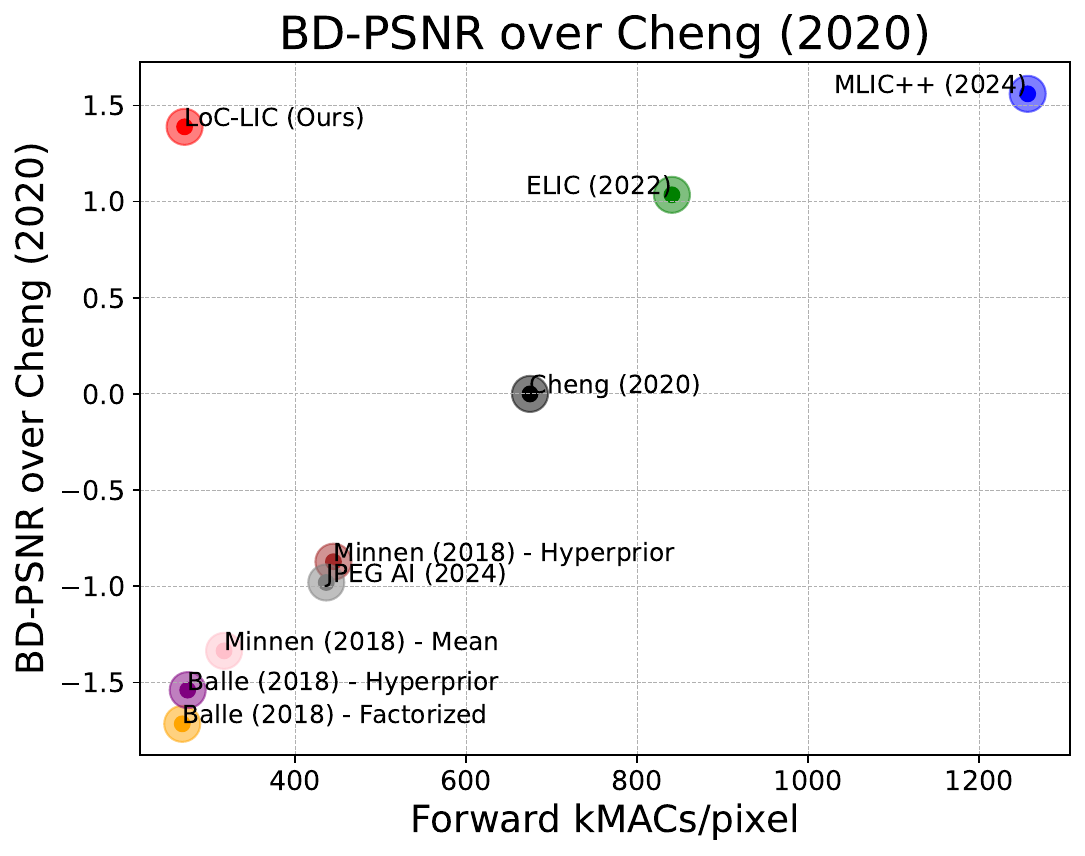}
        \caption{}
        \label{fig:bd_psnr_256_256}
    \end{subfigure}%
    \begin{subfigure}{0.45\textwidth}
        \centering
        \includegraphics[width=\textwidth]{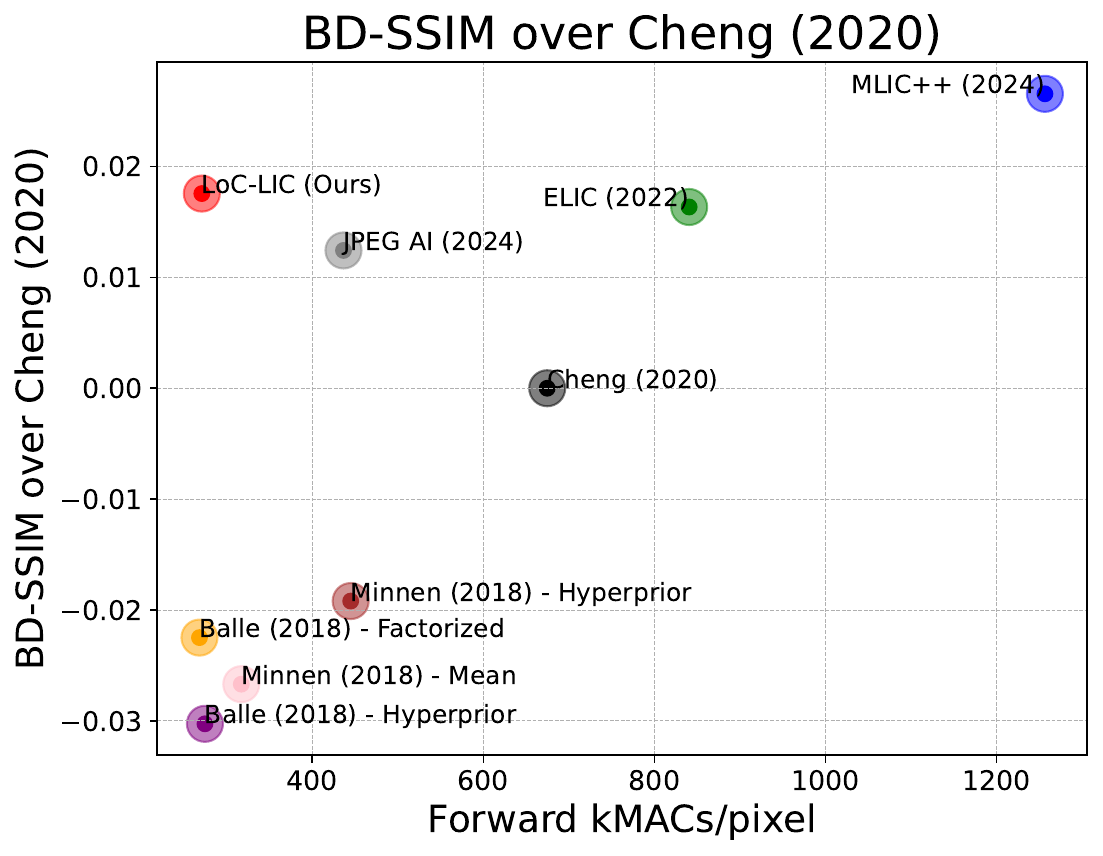}
        \caption{}
        \label{fig:bd_ssim_256_256}
    \end{subfigure}
    \caption{The compression efficiency vs complexity of different learned image compression models. The complexity is measured in terms of (kMAC/Pixel).  (a) BD-PSNR (b) BD-SSIM.}
    \label{fig:complexity_256_256} 
\end{figure*}

\section{Method}
\subsection{Motivation}

Recent advancements in learned image compression models have significantly reduced bit rates. Despite these advancements, the integration of such models into existing systems and devices has been very slow. A primary reason for this slow adoption is the high computational complexity associated with these models, which demand substantial GPU memory and exhibit high operation complexity in terms of multiply-accumulate operations (MACs) per pixel.

In this paper, we introduce an innovative model designed to decrease both computational complexity and GPU memory usage by implementing hierarchical feature extraction from images. Hierarchical feature representation has been successfully applied across various domains, including generative image synthesis, super-resolution imaging, and various medical applications for segmentation and recognition \cite{xuGenerativeHierarchicalFeatures2021,benyahiaMultifeaturesExtractionBased2022,zhuEmotionRecognitionBased2024,wangImageSuperresolutionMethod2024}. However, many of these methods connect basic features from initial layers with more complex features from subsequent layers, leading to increased computational costs. Our approach overcomes this challenge by directly processing composite features while limiting the number of basic features and increasing the number of composite features to achieve computational efficiency.

\subsection{Architecture Overview}

The architecture we propose closely resembles mainstream learned image codecs \cite{jiangMLICLinearComplexity2024,chengLearnedImageCompression2020,minnenJointAutoregressiveHierarchical2018}, as depicted in Figure \ref{fig:model}. The architecture consists of a Hierarchical Feature Encoder that functions as an analysis transform $g_{a_{hf}}$, organizing image data into a latent space hierarchically. This is followed by a quantization function $Q$, which quantizes the latent representation. It also incorporates a hyper-encoder and hyper-decoder featuring a multi-reference entropy model derived from MLIC++ \cite{jiangMLICLinearComplexity2024}, employing both local and global spatial contexts alongside channel and checkerboard attention context models to effectively capture correlations with linear complexity to minimize bit-rate usage.

The reconstruction of the image employs a synthesis transform  $g_{s_{hf}}$. The process can be mathematically formulated as follows:
$$
 y=g_{a_{hf}}(x,\theta), \hat{y} = Q(y), \hat{x} = g_{s_{hf}}(\hat{y},\phi) 
$$
where $x$ represents the input image; $y$ is the unquantized latent vector; $\hat{y}$ is the quantized latent vector; $\theta$ comprises parameters of the analysis transform function; and $\phi$ includes parameters for the synthesis transform function.

\subsection{Hierarchical Feature Transform}

Our hierarchical feature architecture efficiently transforms an input image of height $H$ and width $W$ into progressively deeper feature representations \cite{meyerEfficientLearnedWavelet2024}. Initially, the input image is mapped to basic features with $N$ channels, while height and width are reduced by a factor of 2. In the following layers, features map the number of channels to $2N$ is doubled while their dimensions are reduced by half, effectively enhancing the feature complexity while reducing spatial resolution. This process continues through both encoder and decoder layers.

The mathematical relationship governing this transformation across two sequential layers is expressed as:

$$
\text{out}_{i+1}\left( 2C_{i}, H_{i}/2, W_{i}/2 \right) = \text{in}_{i}(C_{i}, H_{i}, W_{i})
$$

where $\text{out}_{i+1}$ represents the output of layer $i+1$, and $\text{in}_{i}$is the input from layer $i$, characterized by $C_i$ channels with dimensions $H_i \times W_i$.

This architectural design significantly reduces computational complexity. High spatial resolution inputs/feature maps use fewer channels, thereby minimizing computational requirements. Meanwhile, feature maps with a large number of channels have reduced spatial dimensions, thus reducing computational complexity without compromising performance.
\begin{figure}[h]
\centering
\includegraphics[width=0.45\textwidth]{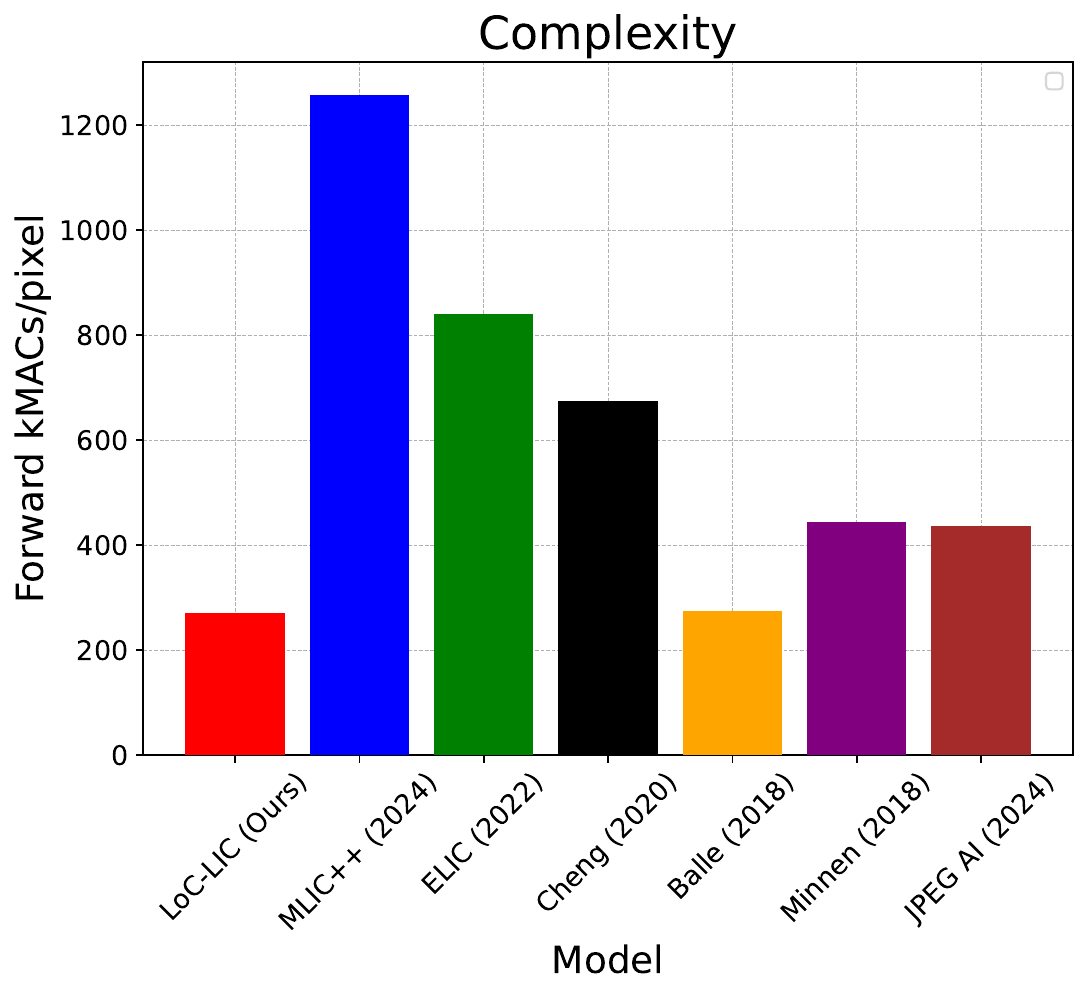}
% \captionsetup{width=0.45\linewidth}
\caption{Comparison of forward complexity between our approach and various learned image compression models}
\label{fig:complexity_256_256_bar}
\end{figure}
\section{Experiments}
 We train our models on $256 \times 256$ randomly cropped images from a custom dataset containing around $10^6$. Our custom dataset images is selected from ImageNet \cite{dengImageNetLargescaleHierarchical2009a}
COCO 2017 \cite{linMicrosoftCOCOCommon2015} Vimeo90K \cite{xueVideoEnhancementTaskOriented2019}, and
DIV2K \cite{agustssonNTIRE2017Challenge2017}. Our objective function consists of two terms. The first one is the mean square error between the original image and the model's output. The second term is the bitrate with a Lagrange multiplier to control the trade-off between the two terms and achieve the target bitrate.

\newcommand{\textwidthplot}{0.33}
\begin{figure*}[h]
    \centering
    \begin{subfigure}{\textwidthplot\textwidth}
        \centering
        \includegraphics[width=\textwidth]{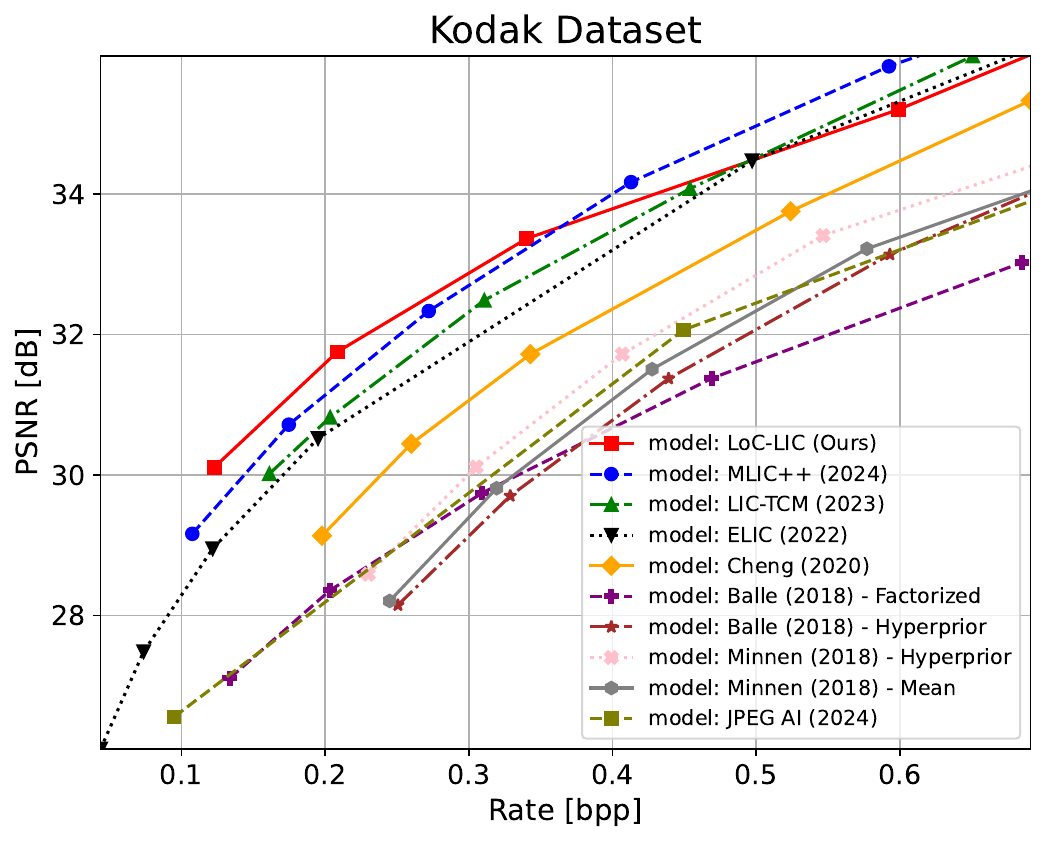}
        \caption{}
        \label{fig:PSNRdBKodakDataset}
    \end{subfigure}%
    ~ % add space between subfigures if needed
    \begin{subfigure}{\textwidthplot\textwidth}
        \centering
        \includegraphics[width=\textwidth]{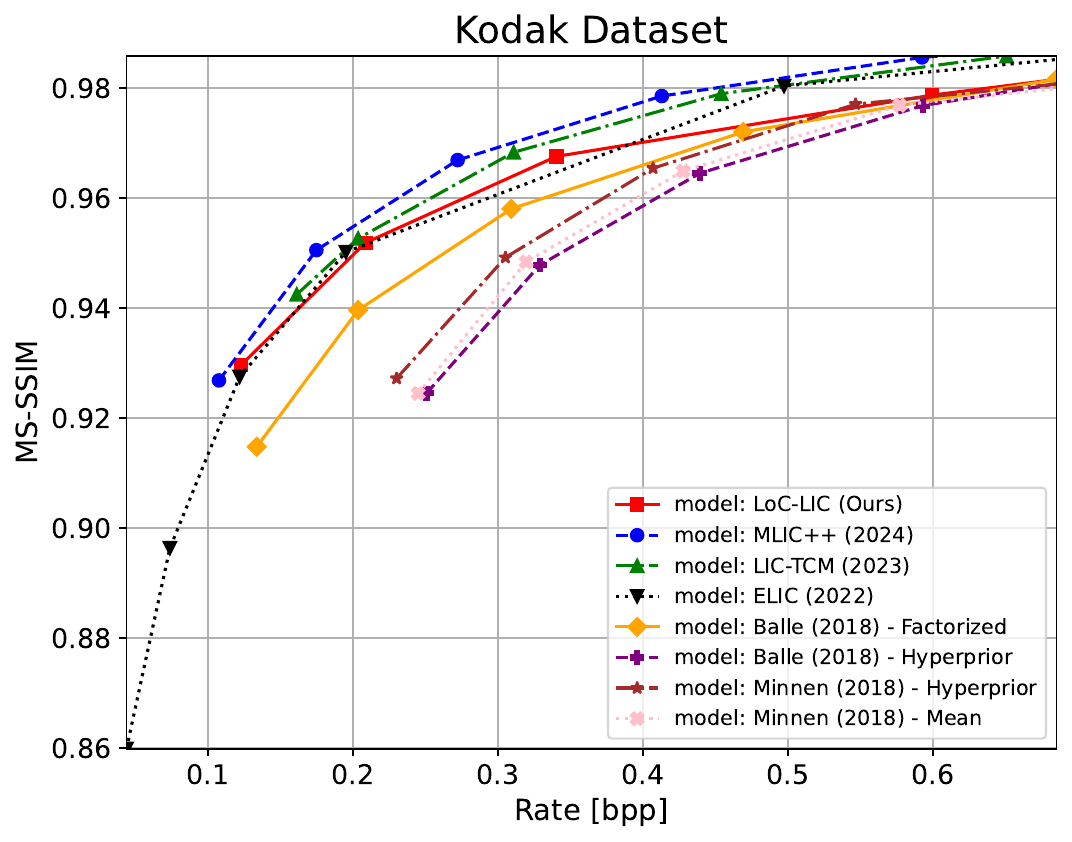}
        \caption{}
        \label{fig:MSSSIMKodakDataset}
    \end{subfigure}%
    ~ % add space between subfigures if needed
    \begin{subfigure}{\textwidthplot\textwidth}
        \centering
        \includegraphics[width=\textwidth]{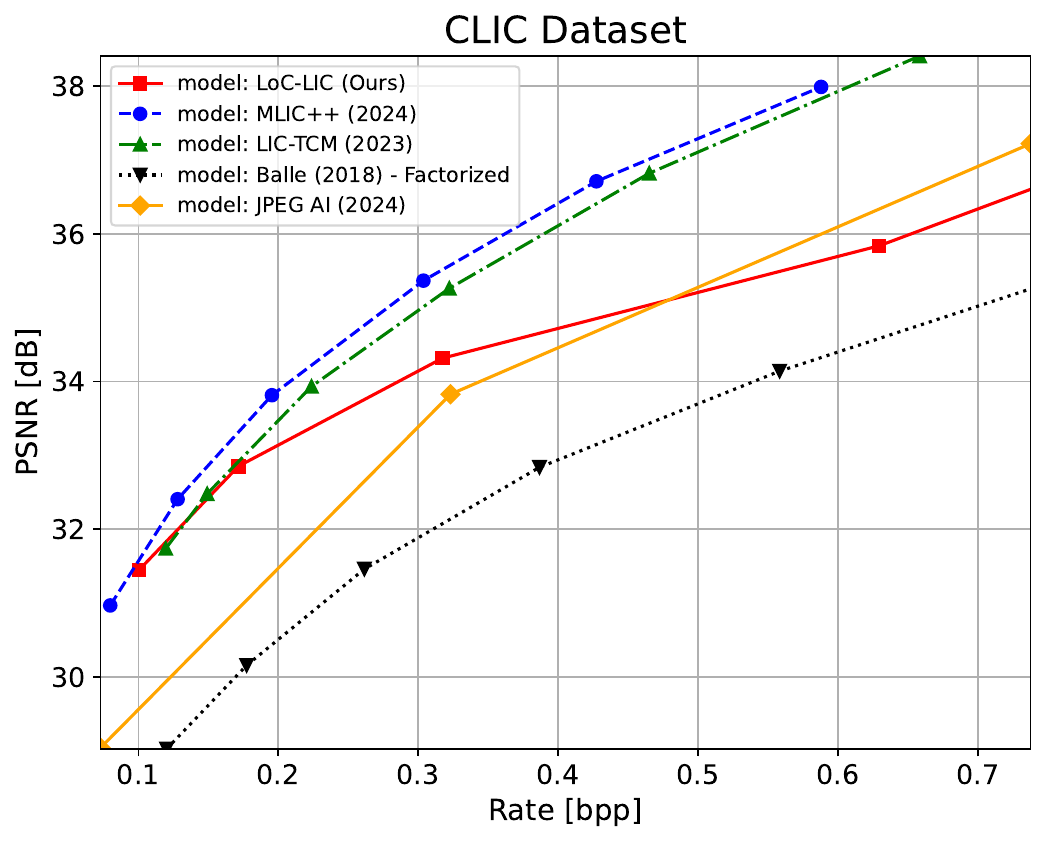}
        \caption{}
        \label{fig:PSNRdBprofessionalvalid2020}
    \end{subfigure}

    \caption{Assessments and comparisons of image compression models using different metrics and datasets. (a) PSNR scores on the Kodak dataset, (b) MS-SSIM scores on the Kodak dataset, and (c) PSNR scores on the CLIC Professional Valid 2020 dataset.}
    \label{fig:combined_assessment_scores} 
\end{figure*}
To assess and compare the performance and generalization capability of our model, we conducted validation experiments on two datasets and evaluated its performance against various models. The first dataset utilized is the Kodak dataset \cite{kodakKodakLosslessTrue1993}, a widely adopted benchmark for validating image compression models comprising 24 images. Additionally, we selected the CLIC Professional Valid 2020 dataset \cite{CLICChallengeLearned}, which contains 41 high-resolution images, making it well-suited for evaluating compression in the current era of digital high-resolution imagery. We compared our approach against several learned image compression models, including MLIC++ \cite{jiangMLICLinearComplexity2024}, LIC-TCM \cite{liuLearnedImageCompression2023}, ELIC \cite{heELICEfficientLearned2022}, JPEG AI \cite{JPEGAI2025} (JPEG-AI-high variant) and two variations of Balle's (2018) model, Factorized and Hyperprior \cite{balleVariationalImageCompression2018}, two variations of Minnen's (2018) model, Mean, and Hyperprior \cite{minnenJointAutoregressiveHierarchical2018}, as well as Cheng's (2020) Anchor model \cite{chengLearnedImageCompression2020}, were included in the comparison. 

\subsection{Quantitative analysis}
We evaluated the complexity of our model in terms of forward operations measured as kMAC/Pixel, comparing this against other models to assess its efficiency. Using Cheng 2020 \cite{chengLearnedImageCompression2020} as a baseline, illustrated in Figure \ref{fig:complexity_256_256} (a), our model exhibited a significantly reduced complexity of approximately 270 kMAC/Pixel while maintaining superior performance over the Cheng 2020 model, which has a complexity of 933 kMAC/Pixel. Moreover, our model outperformed those by Balle (2018) and Minnen (2018) at both lower and higher bit-rates; these models utilize two different approaches corresponding to varying levels of complexity. An average model complexity was considered for comparison purposes. On the other hand, MLIC++ (2024) proved more efficient with a complexity of around 1256 kMAC/Pixel that we could not surpass. Additionally, our model achieves competitive results in terms of the Structural Similarity Index Measure (SSIM) metric, indicating higher values that align with reduced complexity, as shown in Figure \ref{fig:complexity_256_256} (b)

We conducted a comprehensive evaluation of our approach, specifically focusing on a forward path that is responsible for the image encoding and decoding phases. Our analysis, depicted in Figure \ref{fig:complexity_256_256_bar}, compares our method to other advanced learned image compression models. Notably, our model demonstrated the lowest complexity, outperforming leading models such as MILC++.

\newcommand{\textwidthfigure}{0.8}
\begin{figure*}[h]
\centering
\includegraphics[width=\textwidthfigure\textwidth]{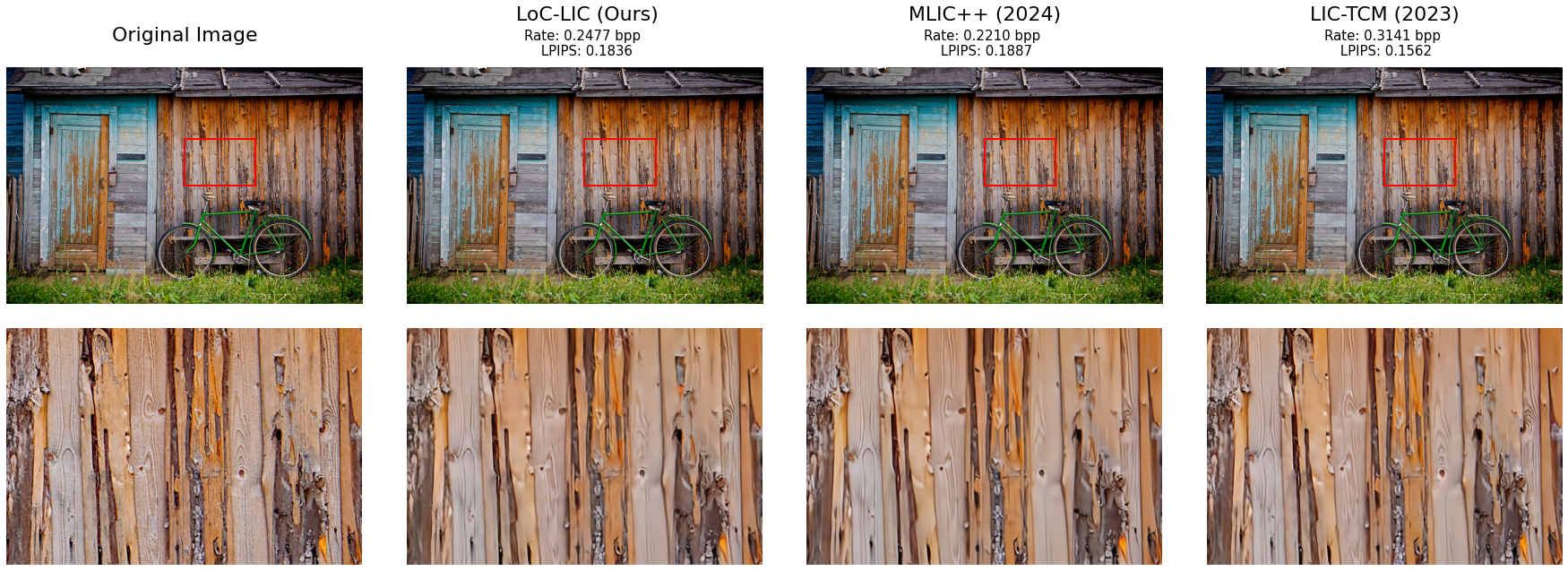}
\captionsetup{width=\textwidthfigure\linewidth}
\caption{Our novel approach performance compared to MLIC++ and LIC-TCM on image num. 3 from the CLIC Professional Valid 2020 dataset.}
\label{fig:CLIC_3_bitrate_zoom_72}
\end{figure*}

\begin{figure*}[h]
\centering
\includegraphics[width=\textwidthfigure\textwidth]{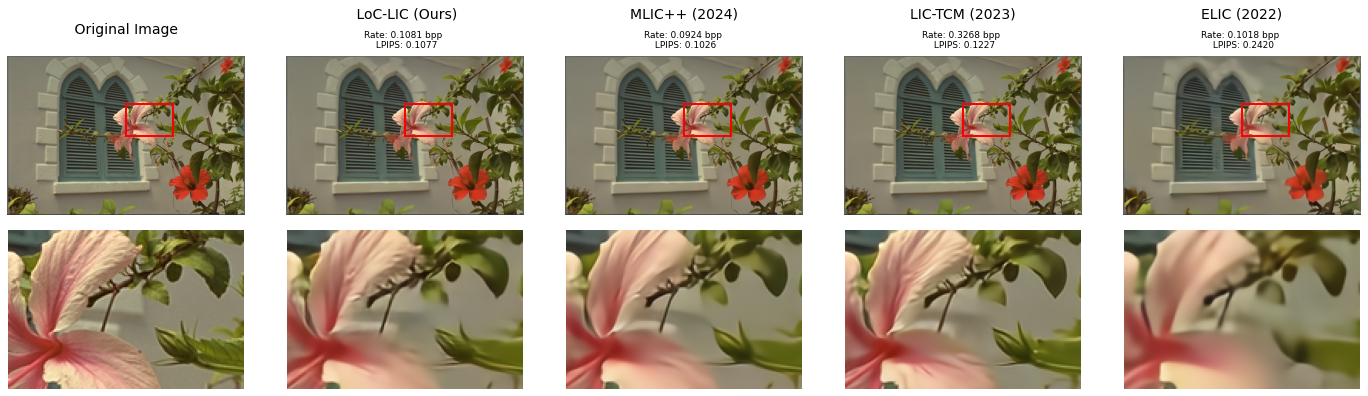}
\captionsetup{width=\textwidthfigure\linewidth}
\caption{Comparison between our approach and different models on image num. 7 from the Kodak dataset.}
\label{fig:Kodak_7_bitrate_zoom_72}
\end{figure*}

We plotted rate-distortion curves for both Peak Signal-to-Noise Ratio (PSNR) and Multi-Scale Structural Similarity Index Measure (MS-SSIM) using the Kodak and CLIC Professional Validation 2020 datasets, as illustrated in Figure \ref{fig:combined_assessment_scores}. Our model exhibits behavior comparable to that of high-complexity models such as MLIC ++ (2024) \cite{jiangMLICLinearComplexity2024} and ELIC (2022), while maintaining significantly lower complexity. In terms of PSNR, our model's performance is shown in Figure \ref{fig:combined_assessment_scores}(a), and for MS-SSIM, the performance is depicted in Figure \ref{fig:combined_assessment_scores}(b). It is observed that our model's performance declines with increasing bit rates in both PSNR and MS-SSIM metrics. For larger images, such as those from the CLIC dataset (Figure \ref{fig:combined_assessment_scores}(c)), our model underperforms compared to other state-of-the-art models. This performance gap can be attributed to training conducted exclusively on $256 \times 256$ pixel images. This limitation could potentially be addressed by including a mixture of $256 \times 256$ and larger sizes, such as $512 \times 512$, during training.

\subsection{Qualitative analysis} 
In our study, we evaluated the visual performance of our model using two distinct datasets, Kodak and CLIC, as illustrated in Figures \ref{fig:CLIC_3_bitrate_zoom_72} and \ref{fig:Kodak_7_bitrate_zoom_72}. Our findings indicate that our model achieves a performance comparable to the state-of-the-art learned image compression model, like MLIC++ (2024) with a marginal increase in the bit rate while maintaining reduced complexity and preserves competitive performance compared to existing models. 

\section{Conclusion}

In this study, we introduce an innovative image compression model with reduced computational complexity, achieving performance comparable to state-of-the-art models. Our method leverages hierarchical feature extraction transforms to significantly lower complexity while effectively maintaining bit rate reduction. We conducted various comparisons with existing learned image compression models, focusing on computational complexity and performance metrics such as PSNR, and LPIPS. Furthermore, we presented our rate-distortion curves with respect to PSNR and MS-SSIM across two benchmark datasets. We also performed qualitative analysis and visual assessments of the compressed images. Our model demonstrated performance on par with state of the art models while retaining minimal complexity.

% \begingroup
% \footnotesize \small  % Change this to any size you prefer, such as \small or \scriptsize
\printbibliography
% \endgroup
\end{document}